\documentclass[aps,prd,twocolumn,amsmath,amssymb,showpacs,superscriptaddress,nofootinbib]{revtex4-1}
\usepackage{amssymb}
\usepackage{graphicx,epsfig}
\usepackage{dcolumn}
\usepackage{bm}
\usepackage{psfig}
\usepackage[caption=false]{subfig}
\usepackage[CJKbookmarks=true,unicode,colorlinks,linkcolor=blue,anchorcolor=blue,citecolor=blue,pdfborder={0 0 0}]{hyperref}
\usepackage{amsfonts}
\usepackage{keyval,graphicx}
\usepackage{textcomp,wasysym}

\begin{document}

\normalsize
\parskip=5pt plus 1pt minus 1pt

\title{Measurements of $\psi^\prime\to\bar{p}K^+\Sigma^0$
       and $\chi_{cJ}\to\bar{p}K^+\Lambda$}

\author{ 
\begin{small}
\begin{center}
M.~Ablikim$^{1}$, M.~N.~Achasov$^{6}$, O.~Albayrak$^{3}$, D.~J.~Ambrose$^{39}$, F.~F.~An$^{1}$, Q.~An$^{40}$, J.~Z.~Bai$^{1}$, Y.~Ban$^{26}$, J.~Becker$^{2}$, J.~V.~Bennett$^{16}$
% ...
\\
\vspace{0.2cm}
(BESIII Collaboration)\\
\vspace{0.2cm} {\it
$^{1}$ Institute of High Energy Physics, Beijing 100049, People's Republic of China\\
$^{2}$ Bochum Ruhr-University, D-44780 Bochum, Germany\\
$^{3}$ Carnegie Mellon University, Pittsburgh, Pennsylvania 15213, USA\\
% ...
$^{6}$ G.I. Budker Institute of Nuclear Physics SB RAS (BINP), Novosibirsk 630090, Russia\\
% ...
$^{16}$ Indiana University, Bloomington, Indiana 47405, USA\\
% ...
$^{26}$ Peking University, Beijing 100871, People's Republic of China\\
% ...
$^{39}$ University of Rochester, Rochester, New York 14627, USA\\
$^{40}$ University of Science and Technology of China, Hefei 230026, People's Republic of China\\
% ...
}\end{center}
\vspace{0.4cm}
\end{small}
}


\begin{abstract}
Using a sample of $1.06\times10^8$ $\psi^\prime$ mesons collected
with the BESIII detector at the BEPCII $e^+e^-$ collider and $\chi_{cJ}$
mesons produced via radiative transitions from the $\psi^\prime$, we report
the first observation for $\psi^\prime\to \bar{p}K^+\Sigma^0
+c.c.$ (charge-conjugate), as well as
improved measurements for the $\chi_{cJ}$ hyperon decays
$\chi_{cJ}\to\bar{p}K^+\Lambda+c.c.$.
The branching fractions are measured to be
$\mathcal{B}(\psi^\prime\to \bar{p}K^+\Sigma^0+c.c)=
(1.67\pm0.13\pm0.12)\times10^{-5}$,
$\mathcal{B}(\chi_{c0}\to\bar{p}K^+\Lambda+c.c.)=
(13.2\pm0.3\pm1.0)\times10^{-4}$,
$\mathcal{B}(\chi_{c1}\to\bar{p}K^+\Lambda+c.c.)=
(4.5\pm0.2\pm0.4)\times10^{-4}$ and
$\mathcal{B}(\chi_{c2}\to\bar{p}K^+\Lambda+c.c)=
(8.4\pm0.3\pm0.6)\times10^{-4}$, where the first error
is statistical, and the second is systematic.
In the decay of $\chi_{c0} \to \bar{p}K^+\Lambda + c.c.$,
an anomalous enhancement near threshold is observed in the invariant
mass distribution of $\bar{p}\Lambda+c.c.$, which cannot be explained
by phase space.
\end{abstract}


\maketitle

\section{Introduction}

The study of hadronic decays of the $c\bar{c}$ states
$J/\psi$, $\psi^\prime$, and $\chi_{cJ}$ could provide
valuable information on perturbative QCD (pQCD) in the 
charmonium-mass regime
and on the structure of charmonia. The color-octet mechanism
(COM), which successfully
described several decay patterns of the P-wave $\chi_{cJ}$
states~\cite{Wong:COM}, may be applicable to other $\chi_{cJ}$ decays. 
Measurements of $\chi_{cJ}$ hadronic decays may provide new input 
into COM and further assist in
understanding the mechanisms of $\chi_{cJ}$ decays. Hadronic
decays of charmonia below the 
$D\bar{D}$ mass threshold are also a good place to search for
previously unknown meson states~\cite{Report:exotic}. The BES
Collaboration has previously reported
observations of near-threshold structures in baryon-antibaryon
invariant-mass distributions in the radiative decay
$J/\psi\to\gamma p\bar{p}$~\cite{BES:ppbar} and the purely hadronic decay
$J/\psi\to p\bar{\Lambda}K^-$ \footnote{Throughout the text,
inclusion of charge conjugate modes is implied if not stated 
otherwise.}~\cite{Bes:pLam}.
It has been suggested theoretically that these states may be
observations of
baryonium~\cite{Bar:Mod}, or caused by final state
interactions~\cite{FSI:Mod}.
Studying the same decay modes in other charmonia may provide
complementary information to improve the knowledge on these unexpected
enhancements. It is also interesting to search for potential structures
formed by $\Lambda\bar{\Lambda}$ and $p\bar{\Sigma}$ pairs, which could assist
in extending the theoretical models.

BESIII has gathered a sample of $1.06\times10^8$ 
$e^+e^-\to\psi^\prime$ events, which
leads to abundant production of $\chi_{cJ}$ states through radiative
decays. This enables us to search for and study the hadronic decays of
the $\chi_{cJ}$ states with high statistics.

\section{Detector}

BEPCII~\cite{NIM:DET} is a double-ring $e^{+}e^{-}$ collider
that has a peak luminosity reaching about $6\times10^{32}$~
$\mathrm{cm}^{-2}\mathrm{s}^{-1}$ at a center of mass energy of $3770$
MeV.
The BESIII~\cite{NIM:DET} detector has a geometrical acceptance
of $93\%$ of $4\pi$ and has four main components:
(1) A small-cell, helium-based ($40\%$ He, $60\%$
C$_{3}$H$_{8}$) main drift chamber (MDC) with $43$ layers providing an
average single-hit resolution of $135$ $\mu$m, and charged-particle
momentum resolution in a $1$ T magnetic field of $0.5\%$ at $1$
GeV$/c$. (2) An electromagnetic calorimeter (EMC) consisting of $6240$
CsI(Tl) crystals in the cylindrical structure barrel and two
endcaps. The energy resolution at $1.0$ GeV is $2.5\%$ ($5\%$) in
the barrel (endcaps), while the position resolution is $6$ mm ($9$ mm)
in the barrel (endcaps). (3) Particle Identification (PID) is provided by a
time-of-flight system (TOF) constructed of $5$-cm-thick plastic
scintillators, with $176$ detectors of $2.4$ m length in two layers in
the barrel and $96$ fan-shaped detectors in the endcaps. The barrel
(endcap) time resolution of $80$ ps ($110$ ps) provides $2\sigma$
$K/\pi$ separation for momenta up to $\sim 1.0$ GeV$/c$.  (4) The muon
system (MUC) consists of $1000$ m$^{2}$ of Resistive Plate Chambers
(RPCs) in nine barrel and eight endcap layers and provides $2$ cm
position resolution.

\section{Monte-Carlo simulation}
Monte-Carlo (MC) simulation of the full detector is used to determine
the detection efficiency of physics processes, optimize event selection
criteria, and estimate backgrounds. The BESIII simulation
program~\cite{SIM:BESIII}
provides an event generator, contains the detector geometry
description, and simulates the detector response and signal
digitization.  Charmonium resonances, such as $J/\psi$ and
$\psi^\prime$, are generated by KKMC~\cite{KKMC:1,KKMC:2}, which
accounts for the effects of initial-state radiation and beam energy spread.
The subsequent charmonium meson decays are produced with
BesEvtGen~\cite{Evtgen:1,Evtgen:2}. The detector geometry and material
description and the transportation of the decay particles through the
detector including interactions are handled by Geant4~\cite{SIM:Geant4}.

\section{Data Analysis}

\subsection{Event selection}

Candidate $\psi^\prime\to\bar{p}K^+\Sigma^0$ and
$\psi^\prime\to\gamma\chi_{cJ}\to\gamma\bar{p}K^+\Lambda$ events,
with $\Sigma^0\to\gamma\Lambda$ and $\Lambda\to p\pi^-$,
are reconstructed using the following selection criteria.

Charged tracks must have their point of closest approach to the
beamline within $\pm 30$ cm of the interaction point in the beam
direction ($|V_z|<30$ cm) and within $15$ cm of the beamline in the plane
perpendicular to the beam ($V_r<15$ cm), and must have the polar angle
satisfying $|\cos\theta|<0.93$. The time-of-flight and energy loss
$dE/dx$ measurements are
combined to calculate PID probabilities for
pion, kaon, and proton/antiproton hypotheses, and each track is assigned
a particle
type corresponding to the hypothesis with the highest confidence level
(C.L.). For this analysis, four tracks identified as 
$p$, $\bar{p}$, $K^+$, and $\pi^{-}$ are required.
To suppress backgrounds from fake tracks, the $\bar{p}$ and $K^+$ are
constrained to the same vertex by vertex fitting, and are required
to satisfy $|V_z|<10$ cm and $V_r<1$ cm in the case of $\gamma \bar{p}
K^+ \Lambda$ modes, and the same procedure is applied for the
respective antiparticle
combinations in the charge-conjugate mode.

Photon candidates are selected in the EMC by requiring a
minimum energy deposition of $25$ MeV within the barrel region
$|\cos\theta|<0.8$, and $50$ MeV within the endcap regions of
$0.86<|\cos\theta|<0.92$.
EMC cluster timing requirements suppress electronic noise and energy
deposits unrelated to the event.

A kinematic fit that enforces momentum and energy conservation (4C)
is applied with the hypothesis $\psi^\prime\to\gamma p\bar{p}K^+\pi^-$,
where the $p$ and $\pi^{-}$ are constrained by $\Lambda$ decay vertex fitting.
For the events with more than one photon candidate,
the combination with the smallest $\chi_{4C}^{2}$ is retained for further
analysis.

\begin{figure*}[htbp]
   \centerline{
   \psfig{file=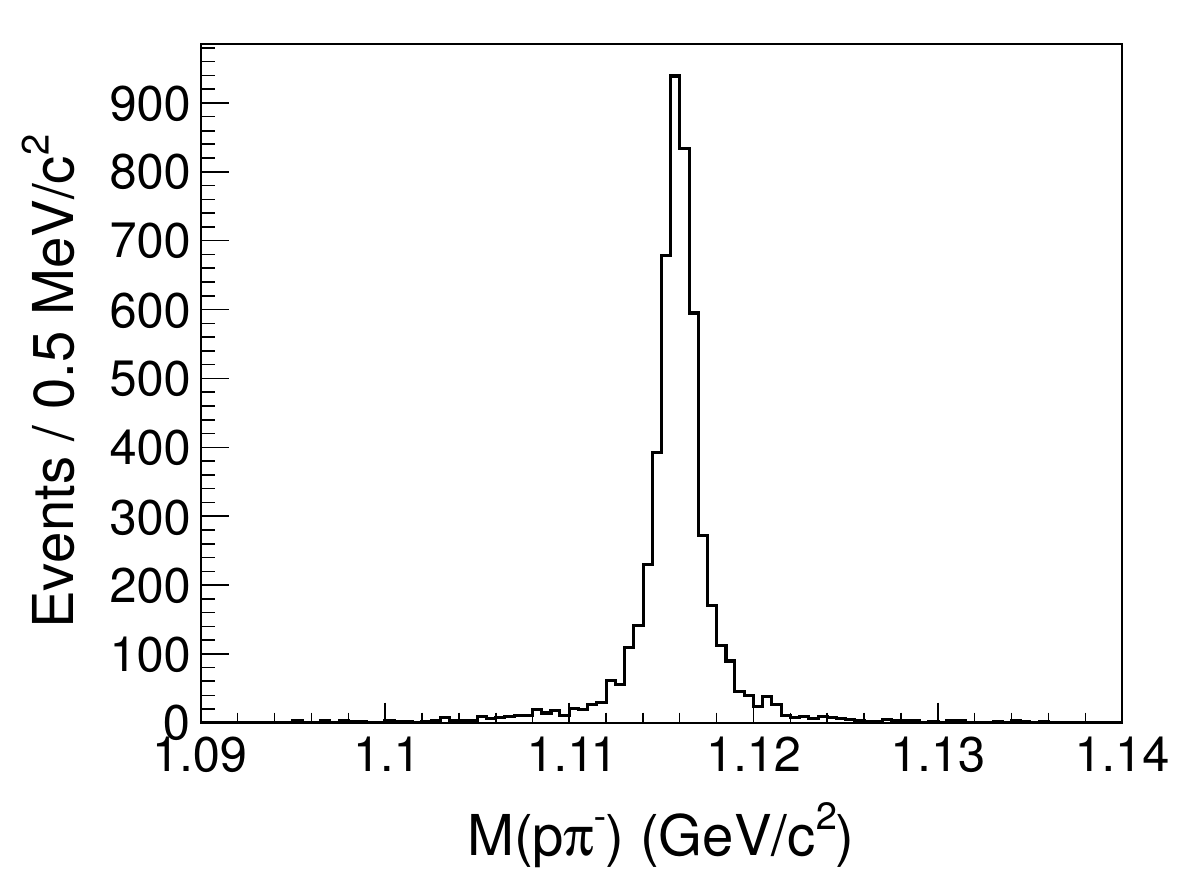,width=7cm,height=6cm, angle=0}
          \put(-107.5,115){\color{red}\vector(0,-1){80}}
          \put(-63.5,115){\color{red}\vector(0,-1){80}}
           \put(-120,110){$\Lambda$}
     }
   \caption{(Color online) 
     The invariant-mass distributions of $p \pi^-$. The vertical (red) 
     arrows show the selection ranges around the $\Lambda$ peak.}
   \label{mass:lambda}
\end{figure*}

\begin{figure*}[htbp]
  \begin{center}
    \subfloat[]{\label{mass:chicj:a}\includegraphics[width=0.47\textwidth]{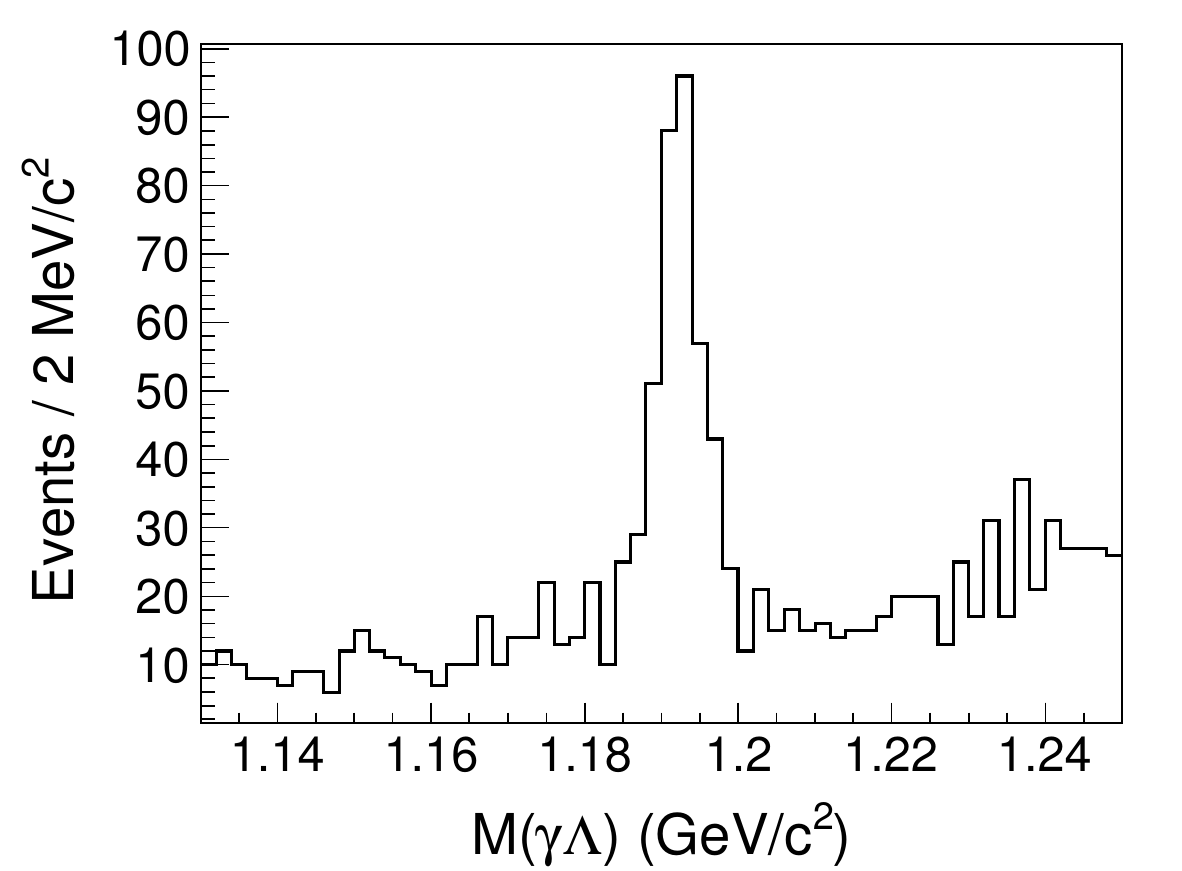}
    \put(-130,110){$\Sigma^{0}$}
    }
    \hfill
    \subfloat[]{\label{mass:chicj:b}\includegraphics[width=0.47\textwidth]{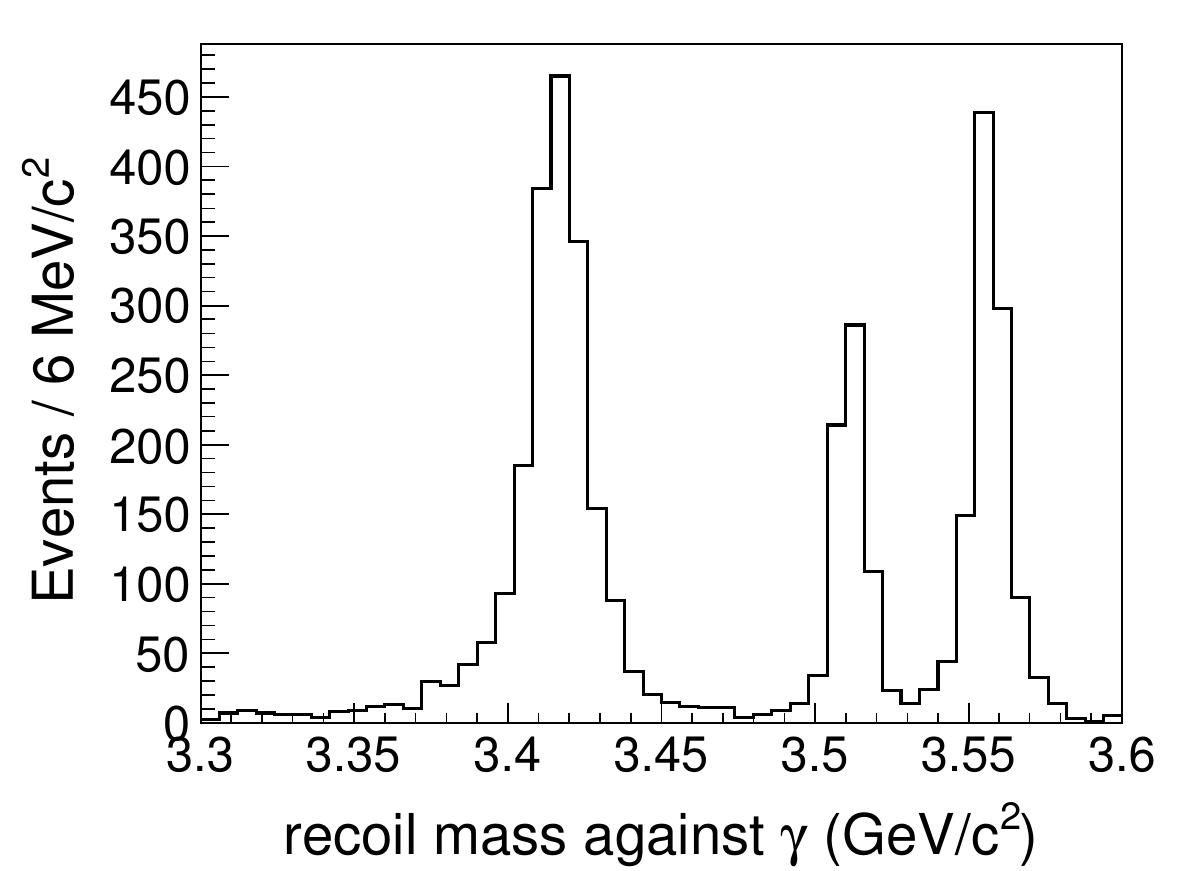}}
    \hfill
    \caption{Distributions of (a) the invariant masses of $\gamma\Lambda$
      and (b) the recoil mass against the $\gamma$ in decays of $\psi^\prime$ after
      vetoing $\psi^\prime\to \bar{p}K^+\Sigma^0$ events.}
    \label{mass:chicj}
  \end{center}
  \end{figure*}

$\Lambda$ candidates are selected by requiring the invariant mass of
$p\pi^-$ to be within $7$ MeV/$c^2$ of the mass of the $\Lambda$ as
given by the PDG~\cite{ref:PDG}, and this distribution is
shown in Figure~\ref{mass:lambda}.
$\Sigma^0$ candidates are formed by
calculating the invariant mass of $\gamma$ and $\Lambda$ candidates,
and this is shown in Figure~\ref{mass:chicj}\nolinebreak\subref{mass:chicj:a}.

After vetoing $\psi^\prime\to \bar{p}K^+\Sigma^0$ events by removing
events where the $\gamma$ and $\Lambda$ have an invariant mass within
$15$ MeV/$c^{2}$ of the $\Sigma^0$ mass~\cite{ref:PDG},
$\chi_{cJ}(J=0,1,2)$ signals
are seen distinctively in the spectrum of recoil mass against the
$\gamma$,
as shown in Figure~\ref{mass:chicj}\nolinebreak\subref{mass:chicj:b}.

\subsection{Background studies}

\begin{figure*}[htbp]
  \begin{center}
    \subfloat[]{\label{Bkg:exl:a}\includegraphics[width=0.47\textwidth]{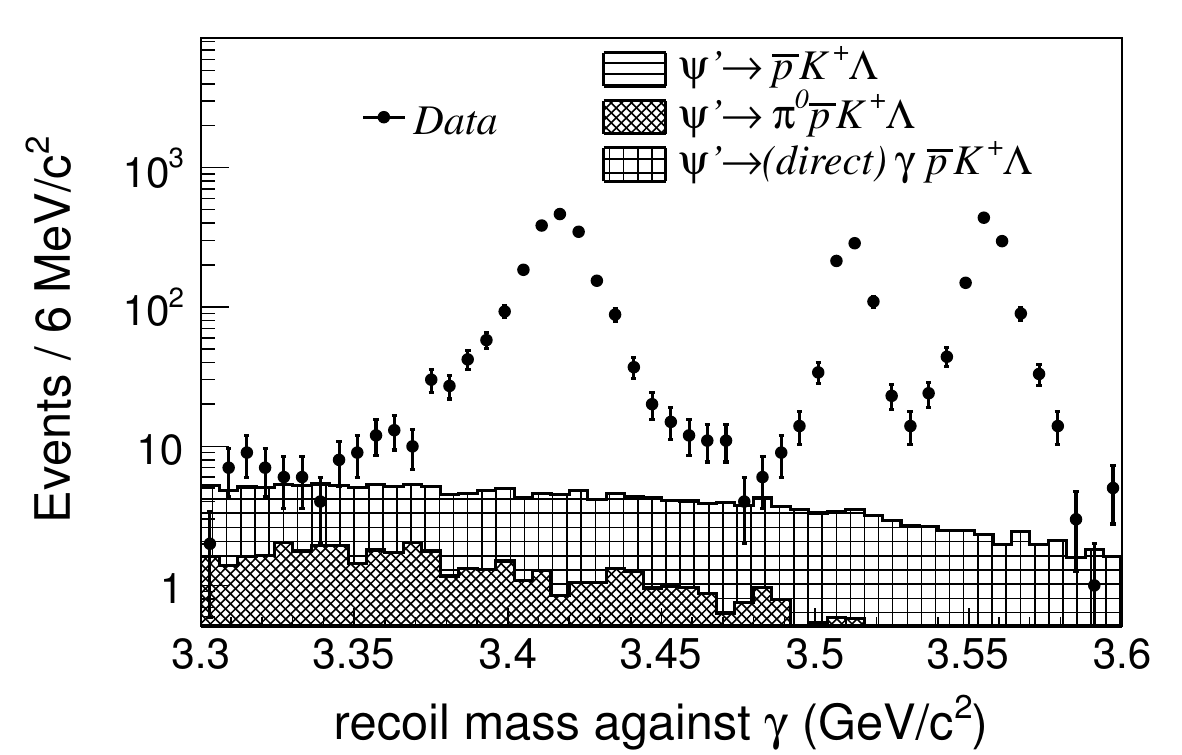}}
    \hfill
    \subfloat[]{\label{Bkg:exl:b}\includegraphics[width=0.47\textwidth]{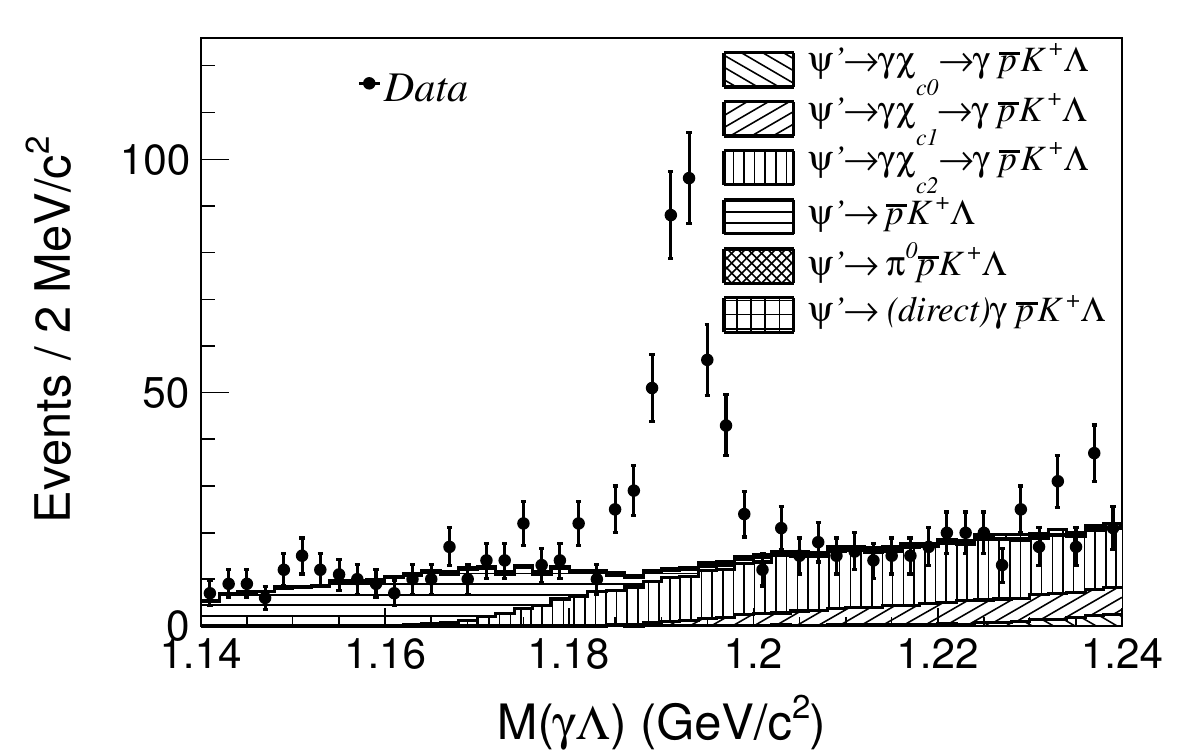}}
    \hfill
   \caption{Comparison of data with exclusive MC samples for
            distributions of (a) the recoil mass against the
            $\gamma$ for $\psi^\prime\to\gamma\chi_{cJ}
	    \to\gamma\bar{p}K^+\Lambda$ and 
	    (b) the $\gamma\Lambda$ invariant mass
	    for $\psi^\prime\to \bar{p}K^+\Sigma^0$.
            The MC samples have been normalized to the total number of
            $\psi^\prime$ events. In figure~(a), the background from
            $\psi^\prime\to \bar{p}K^+\Lambda$ events is too small to 
	    be visible.}
   \label{Bkg:exl}
  \end{center}
\end{figure*}

For the measurements of $\chi_{cJ}\to\bar{p}K^+\Lambda$,
a sample of $1.06\times10^8$ inclusive $\psi^\prime$ MC events are used to
investigate possible backgrounds. The surviving events can be classified
mainly into three decay processes: (1) $\psi^\prime\to \bar{p}K^+\Lambda$,
where a fake $\gamma$ is produced; (2) $\psi^\prime\to\pi^0 \bar{p}K^+\Lambda$
where one $\gamma$ from the $\pi^0$ decay escapes detection; and (3) the direct
decay $\psi^\prime\to\gamma \bar{p}K^+\Lambda$ having the same
final topology with the signal, but not going through an intermediate
$\chi_{cJ}$ state. Accordingly, $2\times10^5$ MC events
for each of the three background processes are produced for further
detailed studies. The same selection criteria are applied to the exclusive
MC samples, and the surviving events are normalized to $1.06\times10^8$
total $\psi^\prime$ MC events. For the normalization procedure, 
the branching fraction $\mathcal{B}=(1.00\pm0.14)\times10^{-4}$ for
$\psi^\prime\to \bar{p}K^+\Lambda$ is quoted  
in the PDG and the other two background modes have branching fractions
in the order of $10^{-5}$,
which we roughly determine from our actual data sample.
Figure~\ref{Bkg:exl}\nolinebreak\subref{Bkg:exl:a} presents the 
distributions of the recoil mass against the $\gamma$ for events that 
survive all cuts for the data and also for these background 
exclusive MC samples.

A similar study is also done for the measurement of $\psi^\prime\to
\bar{p}K^+\Sigma^0$ using the three background modes above together
with $\psi^\prime \to\gamma\chi_{cJ}\to\gamma\bar{p}K^{+}\Lambda\to
\gamma p\bar{p}K^{+}\pi^-$, as shown in Figure~\ref{Bkg:exl}\nolinebreak\subref{Bkg:exl:b}.

In addition, a $42.9$ pb$^{-1}$ data sample, which is approximately a
quarter of the luminosity at $\psi^\prime$ peak, collected at $3.65$ GeV is used to
investigate possible continuum backgrounds. Only $7$ events survived
inside the mass region of $\chi_{cJ}$ for the measurements of
$\chi_{cJ}\to\bar{p}K^+\Lambda$, and are found to be negligible.
For $\psi^\prime\to\bar{p}K^+\Sigma^0$, $110$ events from the continuum
contribution must be subtracted after proper normalization according to
the luminosities.


\subsection{Determination of branching fractions}

\subsubsection{Number of $\psi^\prime\to \bar{p}K^+\Sigma^0$ events}

The decay mode $\psi^\prime\to \bar{p}K^{+}\Sigma^0$ is observed for
the first time, with the main background processes $\psi^\prime\to\gamma
\bar{p}K^{+}\Lambda$, $\psi^\prime\to \pi^0\bar{p}K^{+}\Lambda$,
$\psi^\prime\to \bar{p}K^{+}\Lambda$ and
$\psi^\prime\to\gamma\chi_{cJ}\to\gamma\bar{p}K^{+}\Lambda$.
According to the studies in the previous section, the background shape 
can be described by a linear function,
as shown in Figure~\ref{Bkg:exl}\nolinebreak\subref{Bkg:exl:b}.

\begin{figure*}[htbp]
   \centerline{
    \psfig{file=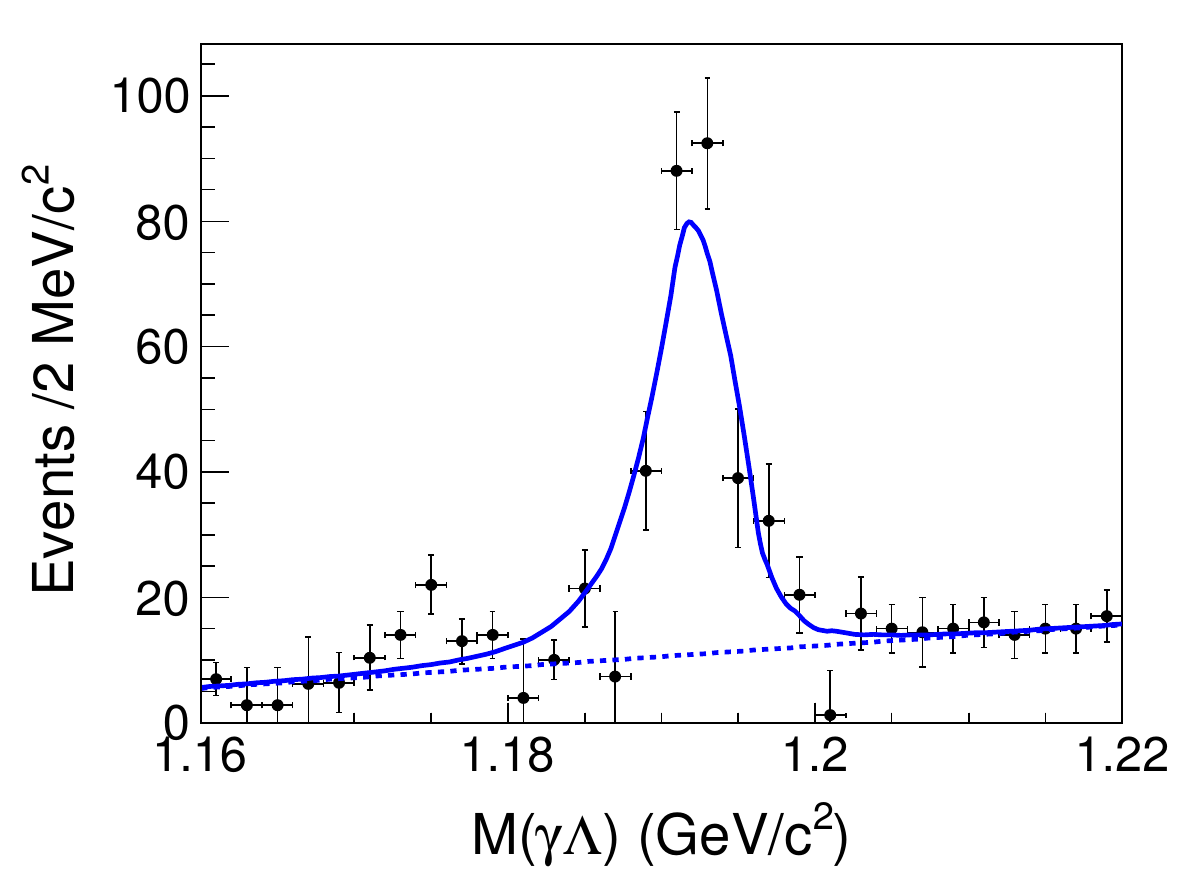,width=7cm,height=6cm, angle=0}
          }
   \caption{(Color online) The shape of the $\Sigma^0$ signal as derived
            from MC simulations which had the mass and width fixed to the PDG
	    values. The fit result is shown by the solid line with a linear
	    background indicated by the dashed line.
            The data points with error bars show the data, where the
            continuum contribution has already been subtracted.
   }
   \label{sigma:fit}
\end{figure*}

A maximum likelihood fit is applied to the spectrum of the
invariant mass of the selected $\gamma$ and $\Lambda$, and we find a
yield of $276\pm21$ events for the $\Sigma^0$ signal. The shape of
the $\Sigma^0$ is obtained
from MC simulation where the mass and width are fixed to the PDG
values. The derived curves are shown in Figure~\ref{sigma:fit}, where
dots with error bars represent the data with continuum contribution
subtracted.

\begin{figure*}[htbp]
    \subfloat[]{\label{sigma:PHSP:a}\includegraphics[width=0.47\textwidth]{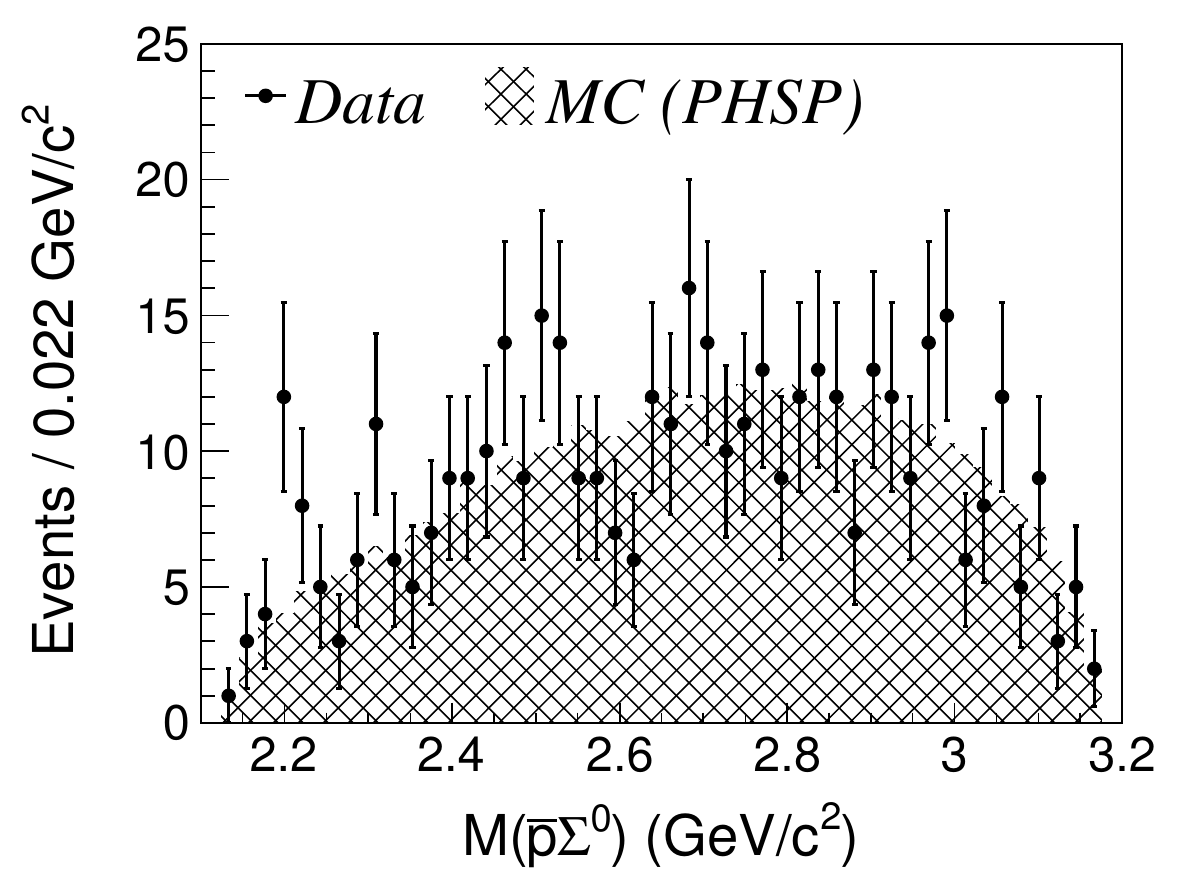}}
    \hfill
    \subfloat[]{\label{sigma:PHSP:b}\includegraphics[width=0.47\textwidth]{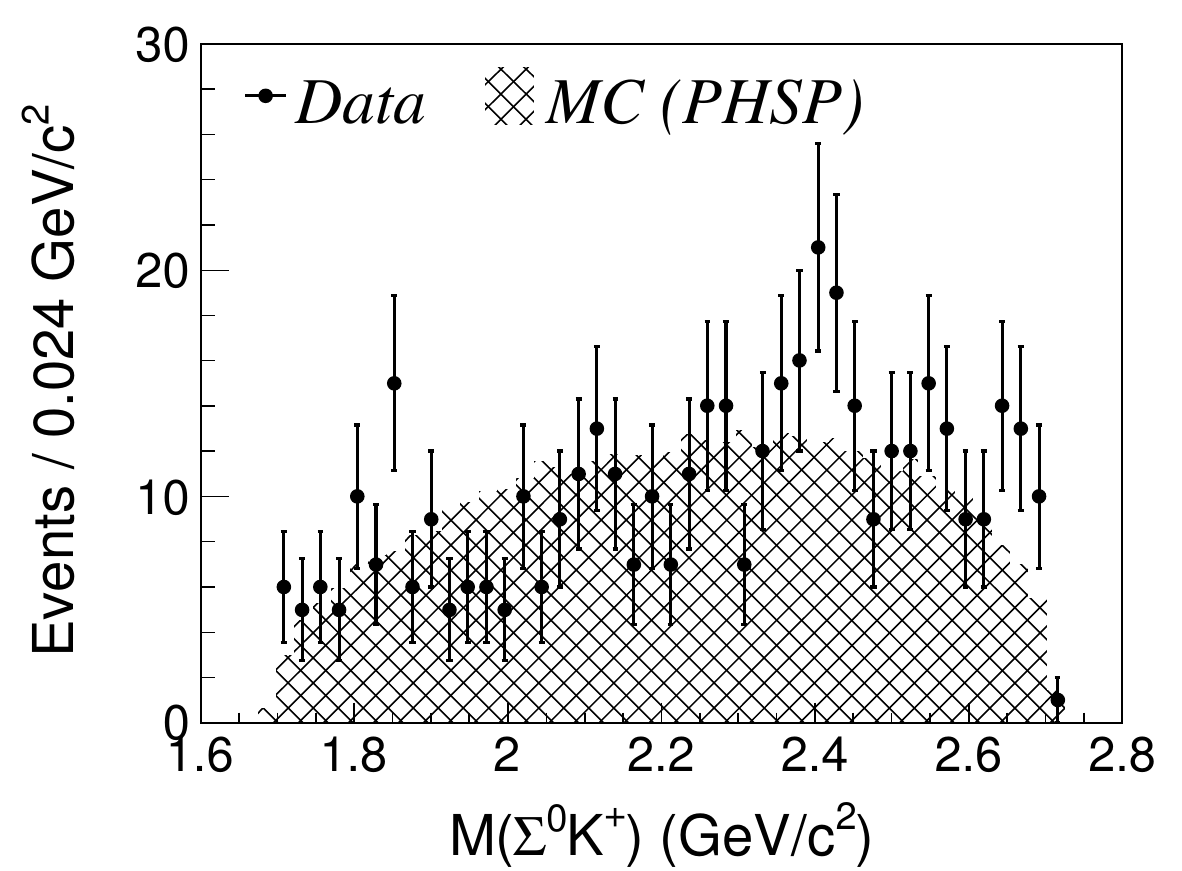}}
    \hfill
   \caption{ Invariant mass spectra of (a) $\bar{p}\Sigma^0$ and
             (b) $\Sigma^0 K^+$ for the reaction $\psi^\prime \to
     \bar{p} K^+ \Sigma^0$. Dots are the data and the hatched regions
             describe MC events generated according to a phase space
             model.
            }
   \label{sigma:PHSP}
\end{figure*}

The detection efficiency for this process is determined to be $24.4\%$
from MC simulation with a phase space model.
The invariant mass spectra of $\bar{p}\Sigma^0$ and $\Sigma^0 K^+$
are shown in Figure~\ref{sigma:PHSP}.

\subsubsection{Number of $\psi^\prime\to\gamma\chi_{cJ}
\to\gamma\bar{p}K^+\Lambda$ events} \label{sec:eff}

\begin{figure*}[htbp]
    \subfloat[]{\label{phsp:data:a}\includegraphics[width=0.47\textwidth]{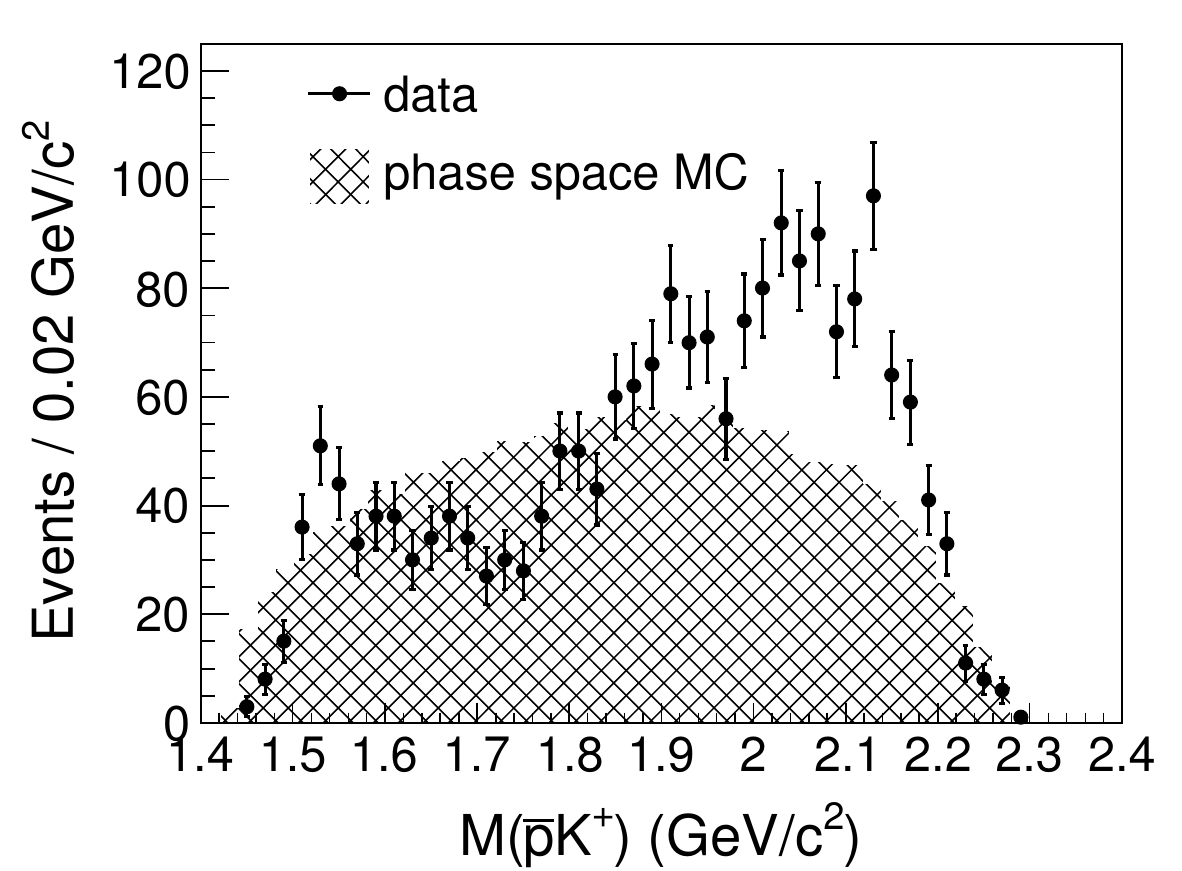}}
    \hfill
    \subfloat[]{\label{phsp:data:b}\includegraphics[width=0.47\textwidth]{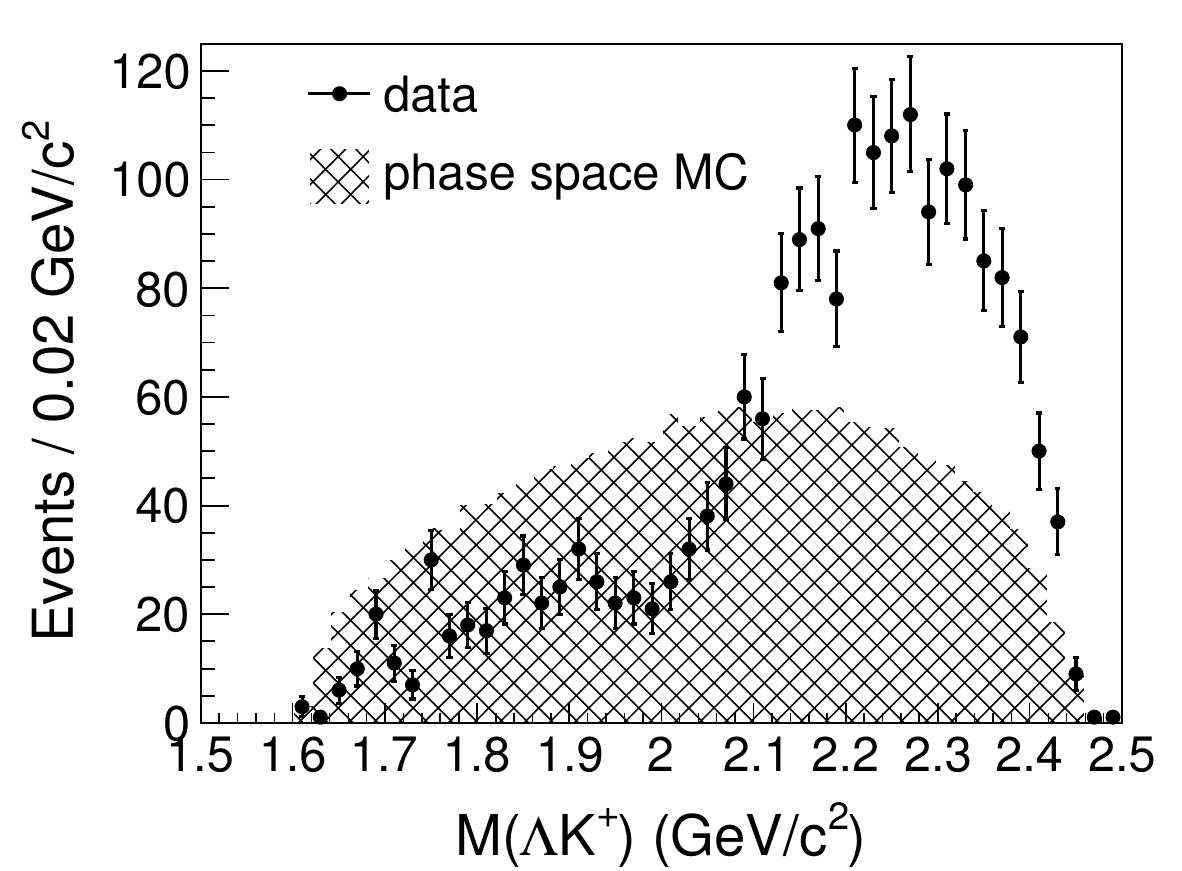}}
    \hfill
    \subfloat[]{\label{phsp:data:c}\includegraphics[width=0.47\textwidth]{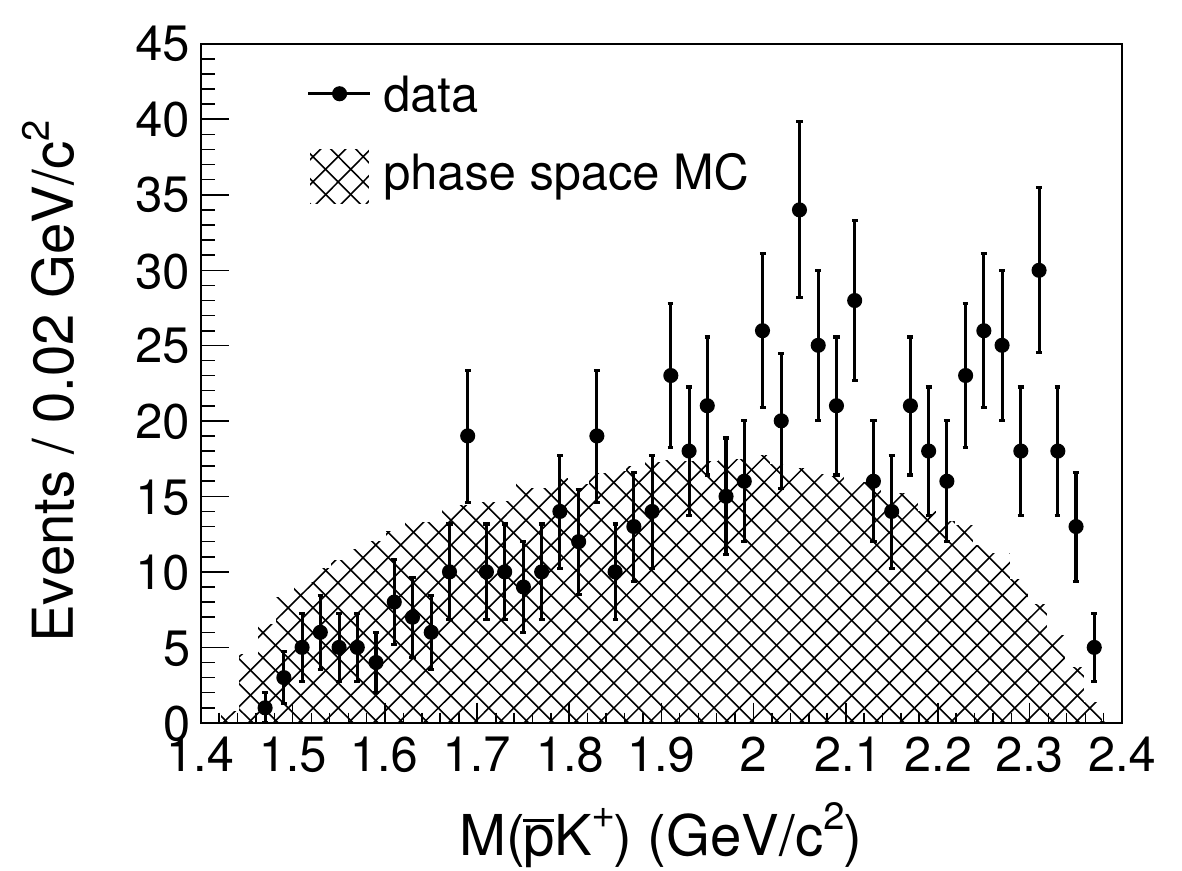}}
    \hfill
    \subfloat[]{\label{phsp:data:d}\includegraphics[width=0.47\textwidth]{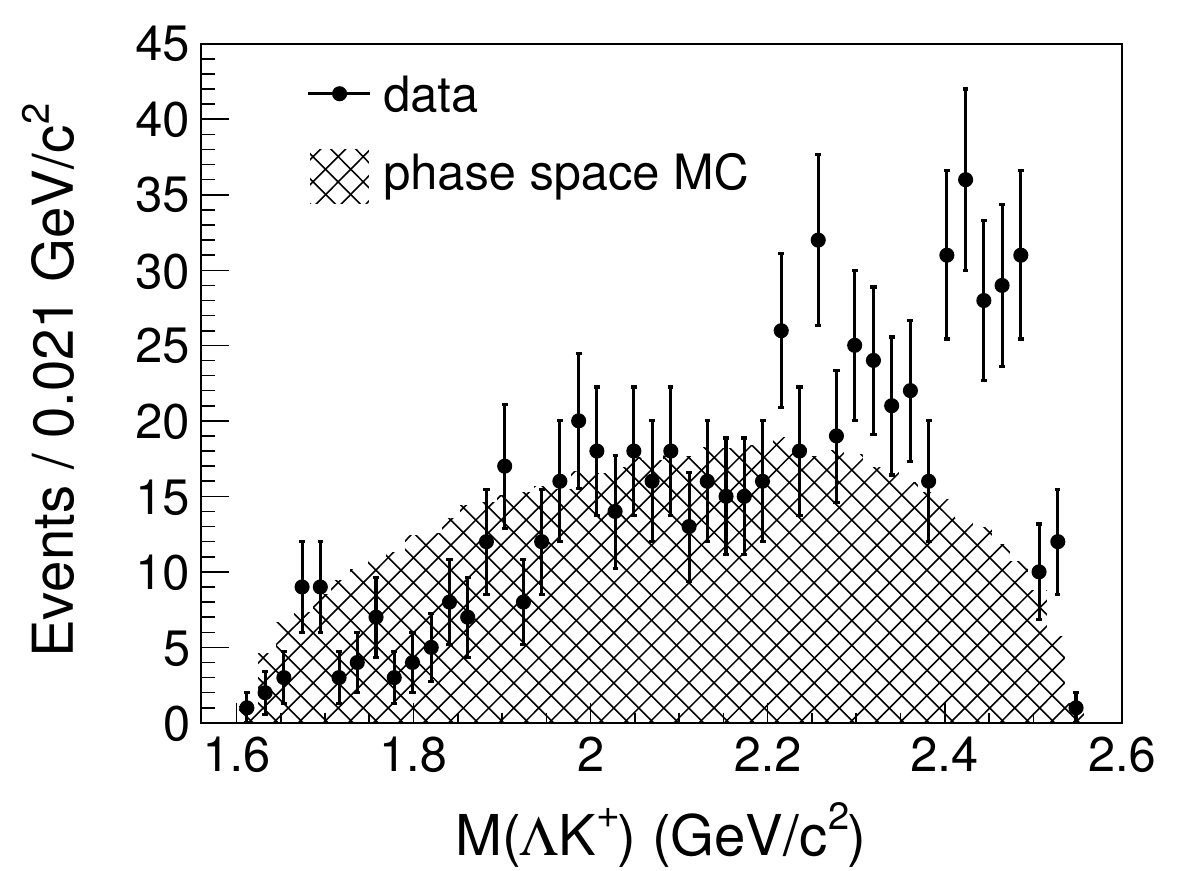}}
    \hfill
    \subfloat[]{\label{phsp:data:e}\includegraphics[width=0.47\textwidth]{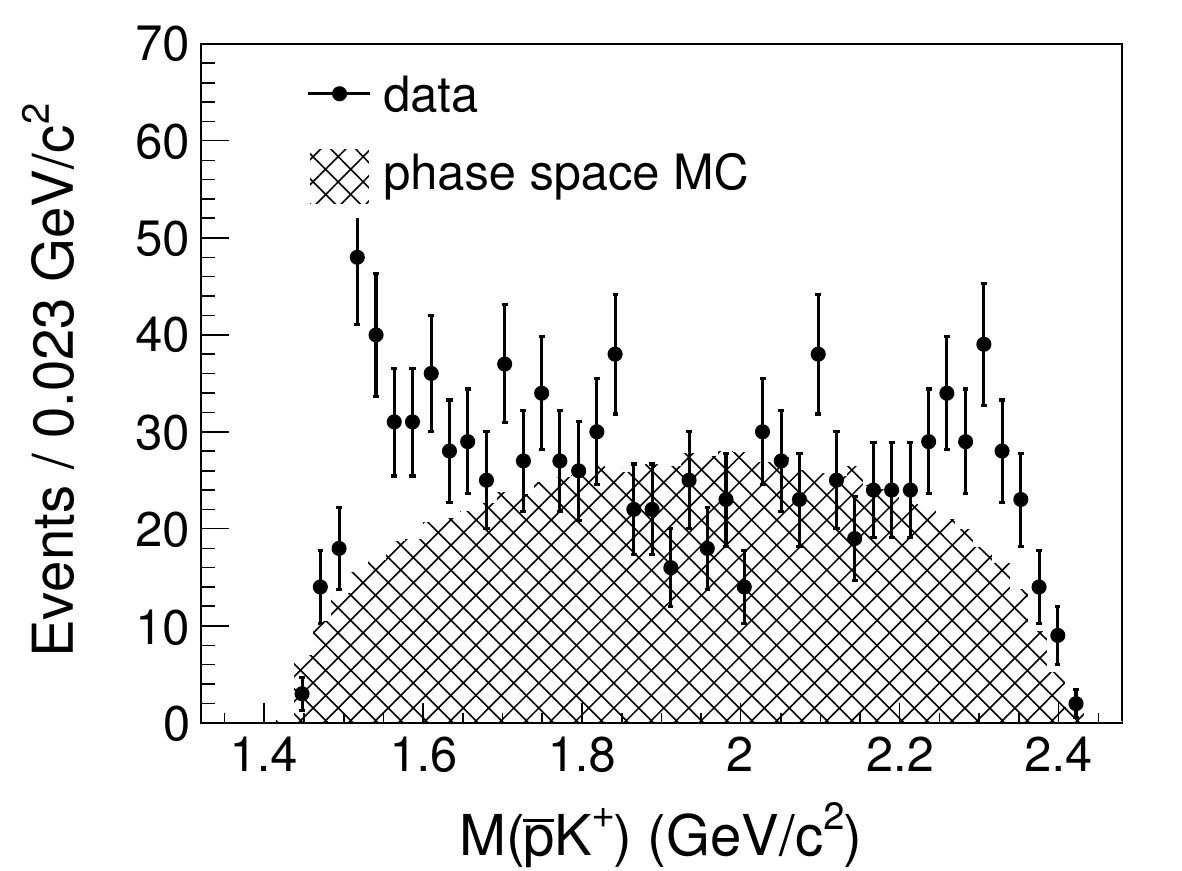}}
    \hfill
    \subfloat[]{\label{phsp:data:f}\includegraphics[width=0.47\textwidth]{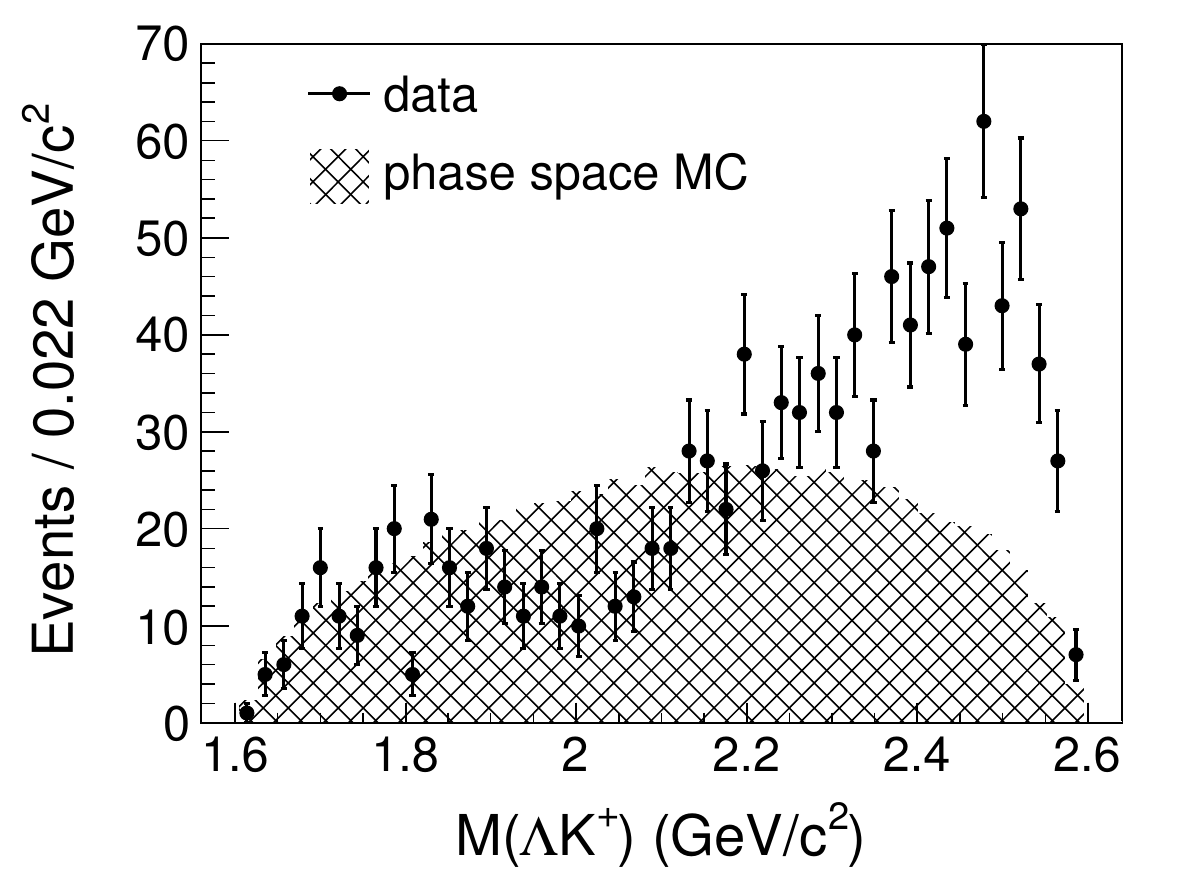}}
    \hfill
   \caption{ Invariant mass spectra of $\bar{p}K^+$ and $\Lambda K^{+}$
     for (a, b) $\chi_{c0}$, (c, d) $\chi_{c1}$ and (e, f) $\chi_{c2}$.
     The dots are the data, and the hatched regions show the
     distribution of MC
     events generated according to a phase space model. Potential
     intermediate states, such as the $\bar{\Lambda}(1520)$
     and $N(1710)$, are seen in the invariant mass
     distributions of $\bar{p}K^+$ and
     $\Lambda K^+$, respectively.
   }
   \label{phsp:data}
\end{figure*}

For the $\chi_{cJ}\to\bar{p}K^+\Lambda$ decays,
obvious inconsistencies exist in the distributions of $\bar{p}K^+$
and $\Lambda K^{+}$ invariant mass between the phase space MC and data,
as shown in  Figure~\ref{phsp:data}, so the detection efficiencies for
the decay modes  $\psi^\prime\to\gamma\chi_{c0, c1,
  c2}\to\gamma\bar{p}K^+\Lambda$  are determined by taking into account
the dynamics of the decay.

For each $\chi_{cJ}$ state, the allowed regions of $M(\bar{p}K^+)$ versus $M(\Lambda K^{+})$
are divided into $25\times25$ areas of equal length ($40$ MeV/$c^2$
for $\chi_{c0}$ and $48$ MeV/$c^2$ for $\chi_{c1}$ and $\chi_{c2}$),
and each area is tagged with an index $ij$.
For each area the number of events $N_\text{data}^{ij}$ for data  
and detection efficiency $\epsilon_{ij}$ are determined individually. Then,
the total number of events ($N_\text{cor}$) is calculated as
$N_\text{cor}=\Sigma_{ij}\frac{N_\text{data}^{ij}}{\epsilon_{ij}}$.
Samples of $5.5\times10^{6}$ MC events are used to determine the
detection efficiencies $\epsilon_{ij}$ of each area for
$\chi_{c0},^{}\chi_{c1},^{}\chi_{c2}$, respectively.

The data belonging to $\chi_{c0}$,
$\chi_{c1}$, and $\chi_{c2}$ are separated using mass windows on
the distribution of recoil mass against the detected $\gamma$ of
$3.35$--$3.48$,
$3.49$--$3.53$, and $3.53$--$3.59$ GeV/$c^2$, respectively.
When extracting $N_\text{data}^{ij}$, the background has been
subtracted using exclusive MC samples according to the results of
background studies. The calculated total numbers of events $N_\text{cor}$
are listed in Table~\ref{num:chicj}.

\begin {table}[htp]
\begin {center}
\caption { The total numbers of events $N_\text{cor}$ for each $\chi_{cJ} \to
           \bar{p}K^+\Lambda$ are derived from
  $N_\text{cor}=\Sigma_{ij}\frac{N_\text{data}^{ij}}{\epsilon_{ij}}$.
  $N_\text{error}$ is the propagated error.} \label{num:chicj}
\begin {tabular}{l>{\hspace{20pt}}c>{\hspace{20pt}}c} \hline\hline
       Modes   &  $N_\text{cor}$  & $N_\text{error}$ \\ \hline
      $\chi_{c0}$ & $8642.7$ & $201.3$ \\
      $\chi_{c1}$ & $2824.0$  & $112.6$ \\
      $\chi_{c2}$ & $4961.0$  & $154.4$ \\\hline\hline

\end {tabular}
\end {center}
\end {table}

\subsubsection{Calculation of branching fractions}

The branching fraction of $\psi^\prime\to \bar{p}K^+\Sigma^0$ is calculated
with
\begin{eqnarray*}
   \mathcal{B} = \frac {N_\text{obs}} {N_{\psi^\prime} \cdot
   \mathcal{B}_{\Sigma^0\to\gamma\Lambda} \cdot
   \mathcal{B}_{\Lambda\to p\pi} \cdot \epsilon },
\end{eqnarray*}
where $N_{\psi^\prime}$ is the total number of $\psi^\prime$ events,
which is measured to be $1.06\times10^{8}$ with an uncertainty of
$0.81\%$~\cite{SYS:number}; the branching fractions $(63.9\pm0.5)\%$
for $\mathcal{B}_{\Lambda\to p\pi}$ and
$100\%$ for $\mathcal{B}_{\Sigma^0\to\gamma\Lambda}$ are taken from
the PDG~\cite{ref:PDG};
$N_\text{obs}$ means the observed number of signals derived from the fit and
$\epsilon$ is the detection efficiency from MC simulation.

The branching fractions for each $\chi_{c0, c1, c2}
\to\bar{p}K^+\Lambda$ are calculated similarly with
\begin{eqnarray*}
   \mathcal{B} = \frac {N_\text{cor}} {N_{\psi^\prime} \cdot
   \mathcal{B}_{\psi^\prime\to\gamma\chi_{cJ}} \cdot
   \mathcal{B}_{\Lambda\to p\pi} },
\end{eqnarray*}
where the branching fractions of the $\chi_{cJ}$ states
($(9.68\pm0.31)\%$, $(9.2\pm0.4)\%$ and $(8.72\pm0.34)\%$ for
$\mathcal{B}(\psi^\prime\to\gamma\chi_{c0})$, $\mathcal{B}(\psi^\prime\to
\gamma\chi_{c1})$ and $\mathcal{B}(\psi^\prime\to\gamma\chi_{c2})$, respectively)
are taken from the PDG~\cite{ref:PDG}.

\subsection{Near-threshold structure}

\begin{figure*}[htbp]
    \subfloat[]{\label{dalitz:plot:a}\includegraphics[width=0.47\textwidth]{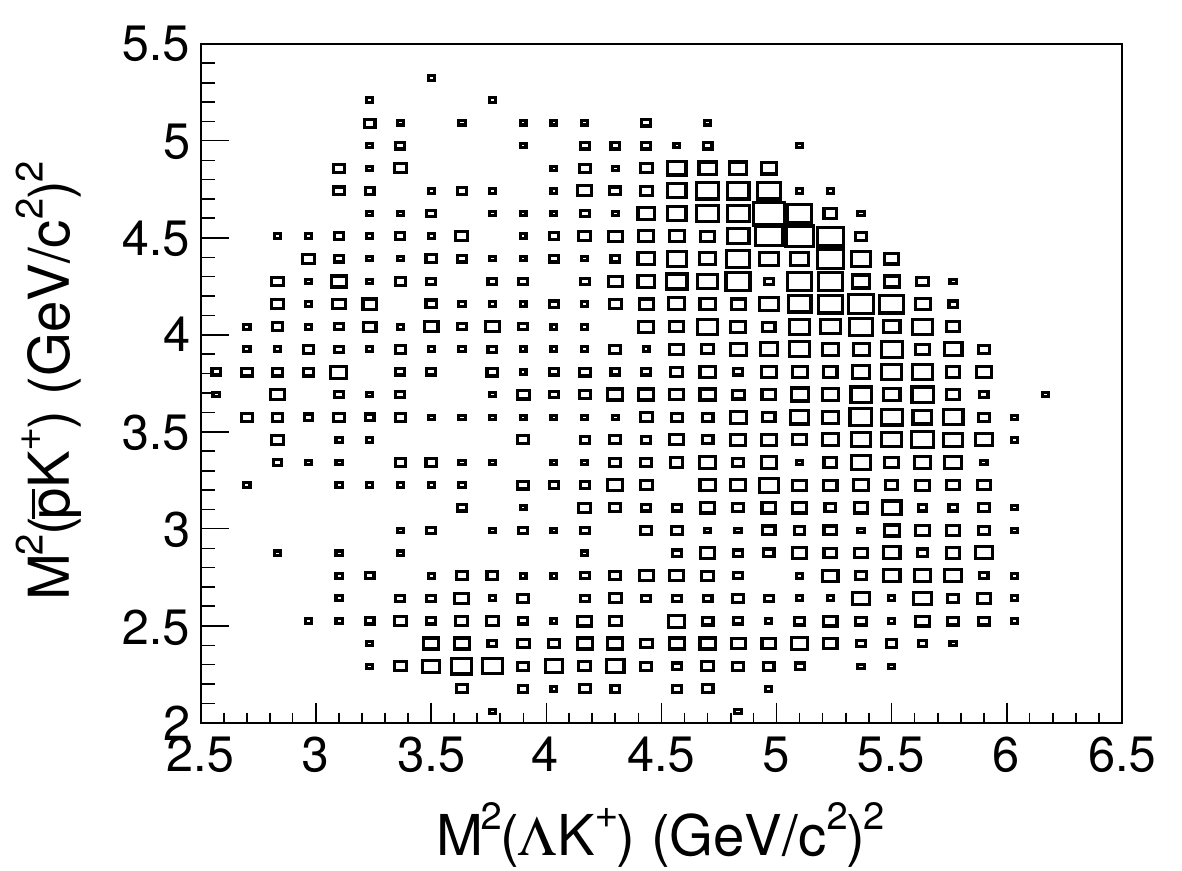}}
    \hfill

    \subfloat[]{\label{dalitz:plot:b}\includegraphics[width=0.47\textwidth]{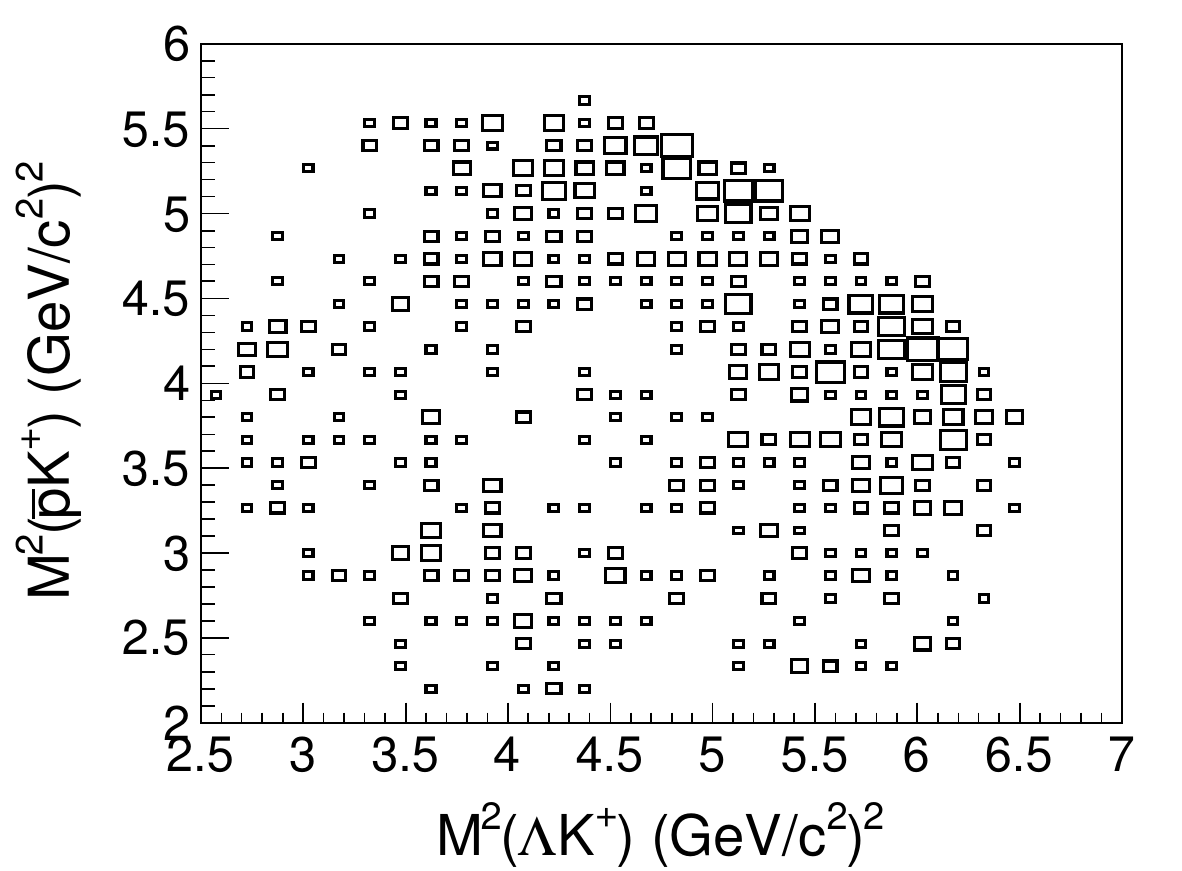}}
    \hfill
    \subfloat[]{\label{dalitz:plot:c}\includegraphics[width=0.47\textwidth]{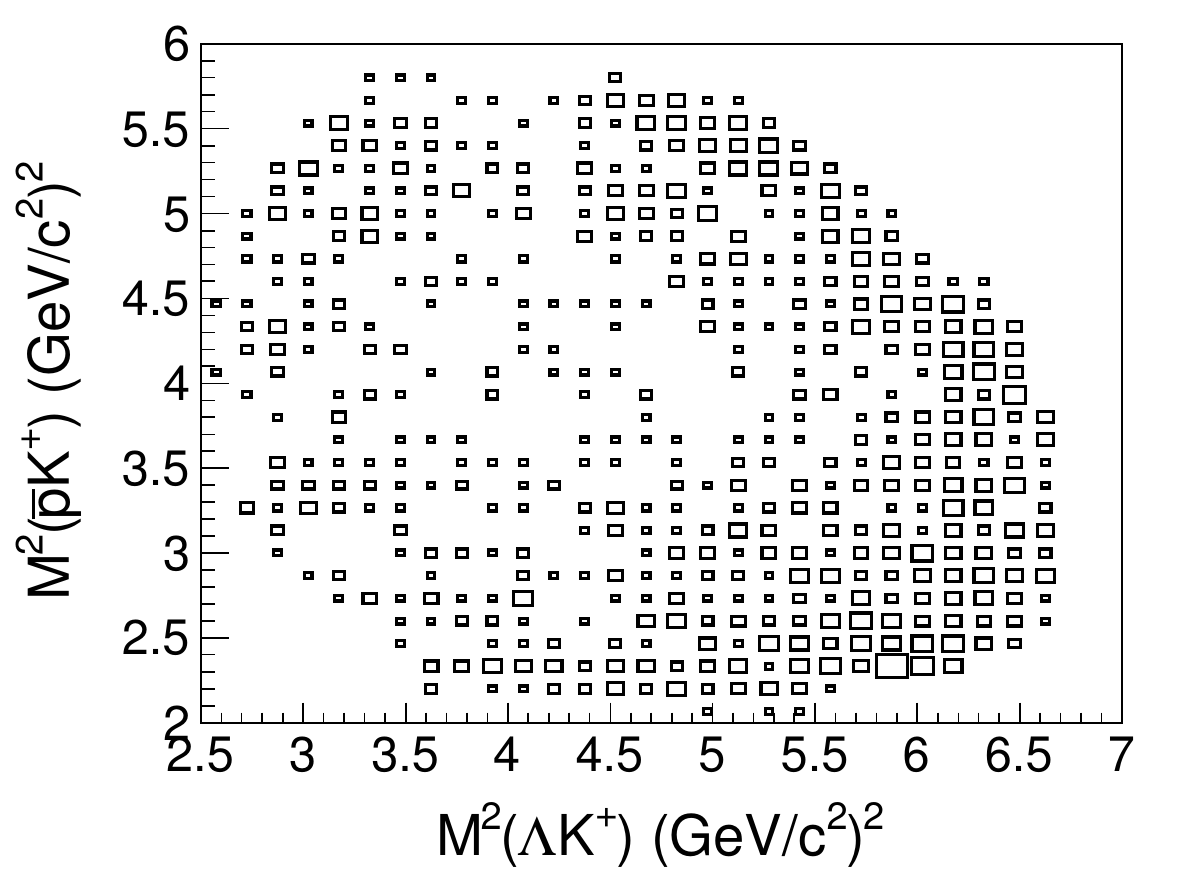}}
    \hfill
   \caption{ Dalitz plots of $M^2(\bar{p}K^+)$ versus $M^2(\Lambda K^+)$
             for (a) $\chi_{c0}$, (b) $\chi_{c1}$ and (c)
             $\chi_{c2}$. A concentration of events in the upper right
             corner shows an enhancement at the $\bar{p} \Lambda$
             threshold.
           }\label{dalitz:plot}
\end{figure*}

The large discrepancies between the data and phase space MC samples
in Figure~\ref{phsp:data} imply that intermediate states
exist in the decays of $\chi_{cJ}\to\bar{p}K^+\Lambda$.
Possible structures are observed in the Dalitz plots shown in
Figure~\ref{dalitz:plot}, and particularly for the $\chi_{c0}$, it
seems that there is a structure in
the near-threshold region of $M(\bar{p}\Lambda)$ reflected by the
anomalous enhancement in the top right corner of the Dalitz plot.

\begin{figure*}[htbp]
    \subfloat[]{\label{mass:plam:a}\includegraphics[width=0.47\textwidth]{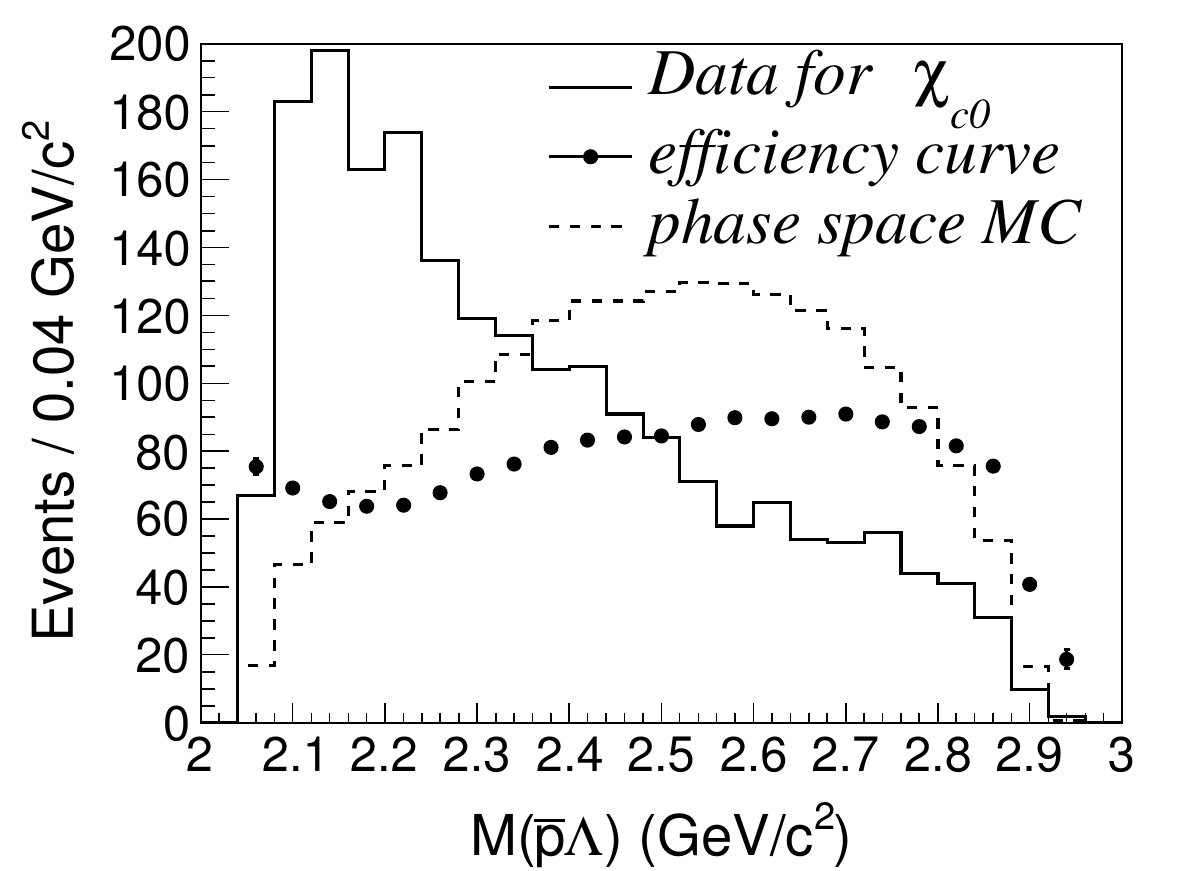}}
    \hfill
    \subfloat[]{\label{mass:plam:b}\includegraphics[width=0.47\textwidth]{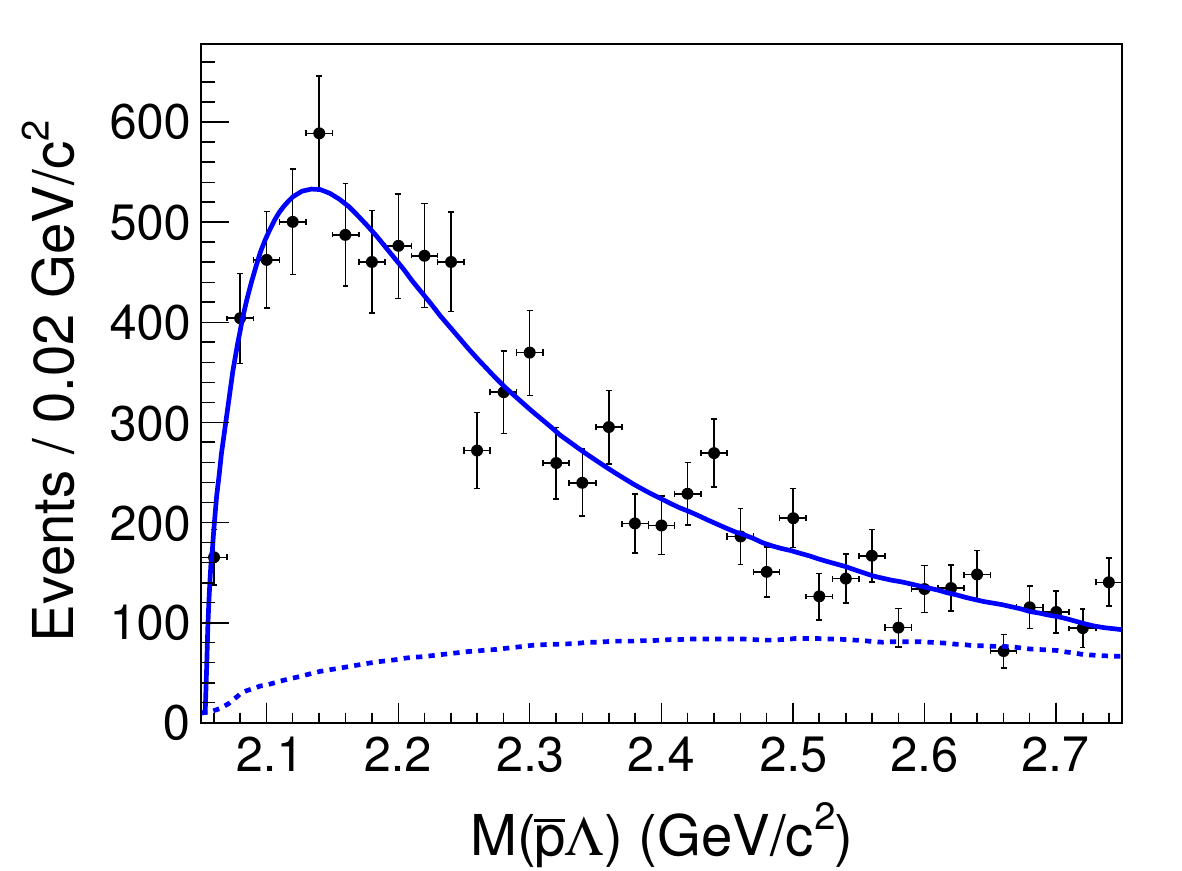}}
    \hfill
   \caption{(Color online)
     (a) Invariant-mass distribution of $\bar{p}\Lambda$ for
     $\chi_{c0}\to\bar{p}K^+\Lambda$, where the dashed line denotes
     the phase space distribution that has been normalized to the
     signal yield. The histogram shows the data and dots present
     the efficiency curve.
     (b) Fit result to a Breit-Wigner function with $J^{P}=0^-$ after
     acceptance correction. The dashed line
     describes the background shape from phase space MC events.
   }\label{mass:plam}
\end{figure*}

Figure~\ref{mass:plam}\nolinebreak\subref{mass:plam:a} shows the 
invariant-mass distribution of
$\bar{p}\Lambda$ for $\chi_{c0}\to \bar{p}K^+\Lambda$, where the
dashed line denotes the phase space distribution that has been normalized to
the signal yield and the dots present efficiencies in each bin. 
Evident discrepancies
are seen near the threshold region. Due to insufficient statistics, in
this analysis a simple fit with a Breit-Wigner function to this region
is done without considering quantum mechanical interference. 
The fit curve for the near-threshold structure is depicted
in Figure~\ref{mass:plam}\nolinebreak\subref{mass:plam:b}, where the distribution
of $M(\bar{p}\Lambda)$
has been corrected by the detector efficiency.
The structure can be fit well with a weighted Breit-Wigner function of
the form
\begin{align}
f(M) \propto
\frac{q^{2L+1}k^{L^\prime+1}}{(M^2-M^2_0)^2-M^2_0\Gamma^2}
\end{align}
where $q$ is the anti-proton momentum in the $\bar{p}\Lambda$ 
rest frame, $k$ is the kaon momentum in the $\chi_{c0}$
rest frame, $L$ ($L^\prime$) denotes the orbital angular momentum 
between the antiproton and $\Lambda$ (between the kaon and $\bar{p}\Lambda$).
On the basis of conservation on $J^{P}$, in the decays of $\chi_{c0}$,
``$L+L^\prime = \text{even~number}$'' can be inferred, and therefore
the only possible spin-parity combinations are $J^{P}=0^-$, $1^+$,
$2^-, \cdots$.
Because the structure is near the $\bar{p} \Lambda$ threshold, 
the relative orbital angular momentum between the antiproton and $\Lambda$ 
is most likely $0$. Therefore, $J^P=0^-$ is used in the fitting process
which gives $M=2.053\pm0.013$ GeV~$/c^2$ and $\Gamma=292\pm14$ MeV for
the Breit-Wigner mass and width parameters.
A shape of the phase space MC is added to describe the
background in the fitting, which is shown as the dashed line
in Figure~\ref{mass:plam}\nolinebreak\subref{mass:plam:b}.

For $\psi^\prime\to\bar{p}K^+\Sigma^0$, the invariant-mass 
spectrum of $M(\bar{p}\Sigma^0)$ was shown in
Figure~\ref{sigma:PHSP}\nolinebreak\subref{sigma:PHSP:a}. In this channel,
there may be similar structures close to the
$\bar{p} \Sigma^0$ threshold, but there is a large
uncertainty due to the relatively small sample size.

\section{Systematic uncertainties}

The main contributions to the systematic uncertainties in the measurements of
the branching fractions originate primarily from the tracking, PID,
photon reconstruction, kinematic fit, branching fractions
of intermediate states, total number of $\psi^\prime$ events, and the
fitting procedure. The results are summarized in Table~\ref{sys:sum}.

The tracking efficiency for MC simulated events is found to agree with
the  data within $1\%$ for each charged track coming from a primary vertex
from analyses of $J/\psi\to K^{\ast}K$ and $J/\psi\to
p\bar{p}\pi^+\pi^-$ events.
For each track from $\Lambda$ (or $\bar{\Lambda}$),
the uncertainty is also $1\%$ according to a study of very clean
$J/\psi\to \bar{p}K^+\Lambda$ events.

The candidates for the selected final states require tracks to be identified as
$p$, $\bar{p}$, $K^{+}$ or $\pi^-$. Comparing data and MC event samples
for $J/\psi\to \bar{p}K^+\Lambda$ and $J/\psi\to K^{\ast}K$, the difference
between MC and data for the particle identification efficiency was found
to be $2\%$ for the antiproton, $1\%$ for the proton and kaon, and
negligible for charged pions.

The difference in the reconstruction efficiency between the data and
MC is about $1\%$ per photon~\cite{SYS:photon}.

To estimate the uncertainty from kinematic fitting, the kinematic fitting 
efficiency is studied using events of
$\psi^\prime\rightarrow\gamma\chi_{c0}\rightarrow\gamma p\bar{p}
\pi^{+}\pi^{-}$ and the difference between data and MC
is found to be $2.8\%$.

Uncertainties due to the mass window requirement for the $\Lambda$ signal
are studied with the control sample $\psi^\prime\to \bar{p}K^+\Lambda$.
The efficiency difference between data and MC is obtained to be $0.4\%$.

Uncertainties in the fitting procedure are obtained by varying
fit intervals and changing the linear background shape to a 
2$^{\text{nd}}$ order Chebyshev polynomial or a MC background shape.
It contributes a $3.3\%$ uncertainty to the 
measurement of $\psi^\prime\to\bar{p}K^+\Sigma^0$.

The uncertainty on the total number of $\psi^\prime$ events was found
to be $0.81\%$ by studying inclusive hadronic $\psi^\prime$
decays~\cite{SYS:number}.

Uncertainties due to the branching fractions of
$\psi^\prime\to\gamma\chi_{cJ}$ are $3.2\%$, $4.3\%$ and $3.9\%$ for
each $\chi_{c0}$, $\chi_{c1}$ and $\chi_{c2}$, respectively~\cite{ref:PDG}.
The uncertainty due to the branching fraction of $\Lambda\to p\pi^-$ is
$0.8\%$~\cite{ref:PDG}.

Uncertainties due to the numbers of areas in the procedure of calculating
total numbers of events for 
$\psi^\prime\to\gamma\chi_{cJ}\to\gamma\bar{p}K^+\Lambda$
are shown as ``2D Binning'' in Table~\ref{sys:sum}.
Detection efficiencies are assumed to be constant within each of these
$25\times25$ sub-areas (see section~\ref{sec:eff}), and as a check,
we varied the number of areas.
Besides the original $25\times25$ binning, three other divisions 
($20\times20$, $30\times30$, $35\times35$) were tried, and the largest
differences among them are taken into account as the systematic
uncertainty due to the binning.

Uncertainties from the mass window requirements of $\chi_{c0}$, $\chi_{c1}$
and $\chi_{c2}$, obtained by changing the $\chi_{cJ}$ selection
window, are shown as item ``Mass Window'' in Table~\ref{sys:sum}, and
are small compared to other errors.

A possible $\Lambda$ polarization in the decays of $\chi_{cJ}$ might
affect detection efficiencies and yield different results. With our
limited statistics, it was not possible to measure the polarization of
the $\Lambda$ in fine bins of the Dalitz plot for each $\chi_{cJ}$
state, but an overall measurement of the $\Lambda$ polarization $P$ was
done for each $\chi_{cJ}$ state that yielded $P=0.04\pm0.07$ for $\chi_{c0}$,
$-0.17\pm0.12$ for $\chi_{c1}$, and $0.22\pm0.09$ for $\chi_{c2}$.
Subsequently, new samples of MC events were then generated with the 
$\Lambda$ having this
polarization $P$, so that the decay distributions are given by
$1+\alpha P \cos\Theta$, where $\Theta$ is the angle between the
$\Lambda$ flight direction
in the $\chi_{cJ}$ rest frame and the $\pi$ direction in the $\Lambda$
rest frame, and $\alpha$ is the weak decay parameter for the $\Lambda$.
The difference in efficiencies with respect to that of phase space MC samples
are taken as a systematic uncertainty.

The total systematic uncertainty is obtained by summing up
uncertainties contributed from all individual sources in quadrature.

\begin {table*}[htpb]
\begin {center}
\caption {Systematic uncertainties in the measurements of the 
           branching fractions in percent ($\%$) } \label{sys:sum}
\begin {tabular}{l|c c c c } \hline\hline
       &  $\psi^\prime\to\bar{p}K^+\Sigma^0$  &
        \multicolumn{3}{p{0.3\textwidth}}{\centering $\chi_{cJ}\to \bar{p}K^+\Lambda$}
        \\ \hline
     &   & \multicolumn{1}{p{0.1\textwidth}}{\centering $\chi_{c0}$ } &
          \multicolumn{1}{p{0.1\textwidth}}{\centering $\chi_{c1}$ }    &
        \multicolumn{1}{p{0.1\textwidth}}{\centering $\chi_{c2}$ }
         \\\hline
     Tracking  & $4.0$ & $4.0$ & $4.0$ & $4.0$      \\
     PID  &  $4.0$ & $4.0$ & $4.0$ & $4.0$ \\
     Photon Recon. & $1.0$ & $1.0$ & $1.0$ & $1.0$ \\
     Kinematic Fit & $2.8$ & $2.8$ & $2.8$ & $2.8$ \\
     Fitting & $3.3$ & $ --- $ & $ --- $ & $ --- $ \\
     $\Lambda$ mass window & $0.4$ & $0.4$ & $0.4$ & $0.4$ \\
     Intermediate states & $0.8$ & $3.3$ & $4.4$ & $4.0$ \\
     $N_{\psi^\prime}$ &   $0.81$ & $0.81$ & $0.81$ & $0.81$ \\
     2D Binning & $ --- $ & $1.3$ & $0.7$ & $1.1$   \\
     Mass Window & $ --- $ & $<0.1$ & $0.7$ & $0.4$     \\
     $\Lambda$ Polarization & $ --- $ & $1.3$ & $0.4$ & $1.8$    \\\hline
     Total & $7.3$ & $7.5$ & $7.9$ & $7.6$  \\\hline \hline
\end {tabular}
\end {center}
\end {table*}

\section{RESULTS AND DISCUSSION}

We observe the decay mode $\psi^\prime\to\bar{p}K^+\Sigma^0+c.c.$
for the first time
and improve the measurements for the decays of
$\chi_{cJ}\to \bar{p}K^+\Lambda+c.c.$,
using $1.06\times10^8$ $\psi^\prime$ events collected with BESIII
detector at the BEPCII collider. The branching fractions are
listed in Table~\ref{tab:sum}.

\begin {table*}[htp]
\begin {center}
\caption { The branching fractions for $\psi^\prime\to\bar{p}K^{+}\Sigma^0+c.c.$
           and $\chi_{cJ}\to \bar{p}K^{+}\Lambda+c.c.$,
           where the first errors are statistical and second ones systematic.
         } \label{tab:sum}
\begin {tabular}{l>{\hspace{2pt}}c>{\hspace{2pt}}c>{\hspace{2pt}}c
                 >{\hspace{2pt}}c} \hline\hline
         channel     & $\psi^\prime\to\bar{p}K^{+}\Sigma^0+c.c.$ &
                $\chi_{c0}\to\bar{p}K^{+}\Lambda+c.c.$  &
                  $\chi_{c1}\to\bar{p}K^{+}\Lambda+c.c.$  &
                  $\chi_{c2}\to\bar{p}K^{+}\Lambda+c.c.$  \\ \hline
     $\mathcal{B}$(BESIII) \hspace{0.5cm} & $(1.67\pm0.13\pm0.12)\times10^{-5}$  &
                     $(13.2\pm0.3\pm1.0)\times10^{-4}$ & $(4.5\pm0.2\pm0.4)\times10^{-4}$
                   & $(8.4\pm0.3\pm0.6)\times10^{-4}$  \\ 
      PDG   &   & $(10.2\pm1.9)\times10^{-4}$ & $(3.2\pm1.0)\times10^{-4}$  
                & $(9.1\pm1.8)\times10^{-4}$  \\\hline\hline 
\end {tabular}
\end {center}
\end {table*}

For the $\bar{p}K^+\Lambda+c.c.$ final state in the decays of $\chi_{c0}$,
an anomalous enhancement is observed in the invariant-mass distribution of
$\bar{p}\Lambda+c.c.$, which could correspond to the structure
observed in the decay $J/\psi\to
p\bar{\Lambda}K^-$~\cite{Bes:pLam}. It is of great interest
that the structure is located very close to the mass threshold of
$\bar{p}\Lambda+c.c.$, and this may be accounted for as a 
quasibound dibaryon state or as an
enhancement due to a final-state interaction, or simply as an interference 
effect of high-mass $N^{\ast}$ and $\Lambda^{\ast}$. 
Our new measurements may aid in the theory of
charmonia decays, and also be a guide in the calculation of decay
modes into strangeness dibaryon systems.
A detail study on the near-threshold structure is expected with larger 
statistics in future BESIII running.

\section*{Acknowledgments}
The BESIII collaboration thanks the staff of BEPCII and the computing 
center for their hard efforts. This work is supported in part by the 
Ministry of Science and Technology of China under Contract 
No. 2009CB825200; National Natural Science Foundation of China (NSFC) 
under Contracts Nos. 10625524, 10821063, 10825524, 10835001, 
10935007, 11005109, 11079030, 11125525, 11179007, 11275189; 
Joint Funds of the National Natural Science 
Foundation of China under Contracts Nos. 11079008, 11179007; 
the Chinese Academy of Sciences (CAS) Large-Scale Scientific Facility 
Program; CAS under Contracts Nos. KJCX2-YW-N29, KJCX2-YW-N45; 100 
Talents Program of CAS; Research Fund for the Doctoral Program of Higher Education 
of China under Contract No. 20093402120022; 
German Research Foundation DFG under Contract 
No. Collaborative Research Center CRC-1044; Istituto Nazionale di Fisica 
Nucleare, Italy; Ministry of Development of Turkey under Contract 
No. DPT2006K-120470; U. S. Department of Energy under Contracts 
Nos. DE-FG02-04ER41291, DE-FG02-94ER40823; U.S. National Science 
Foundation; University of Groningen (RuG) and the Helmholtzzentrum 
fuer Schwerionenforschung GmbH (GSI), Darmstadt; WCU Program of 
National Research Foundation of Korea under Contract No. R32-2008-000-10155-0

\end{document}